\documentclass[12pt]{article}
\usepackage[T2A]{fontenc}
\usepackage[utf8]{inputenc}
\usepackage{setspace,amsmath}
\usepackage{bbold}
\usepackage{amsfonts}
\usepackage{amssymb}
\usepackage{amsthm}
\usepackage{cite}
\usepackage{fontenc}
\usepackage{feynmp-auto}
\usepackage[dvips]{graphicx}
\usepackage[unicode,linktocpage=true,plainpages=false,pdfpagelabels=false]{hyperref}
\newtheorem{theorem}{Theorem}[section]

\newtheorem{lemma}[theorem]{Lemma}
\begin{document}
\title{About application of the matrix formalism of 
the heat\\ kernel to number theory}
\author{A.~V.~Ivanov\thanks{E-mail:regul1@mail.ru}\\
{\it St. Petersburg Department of Steklov Mathematical 
Institute of}\\ {\it Russian Academy of Sciences, 27 Fontanka, 
St. Petersburg, Russia}}
\date{\vskip 15mm}
\maketitle
\begin{abstract}
Earlier in the study of the combinatorial properties 
of the heat kernel of Laplace operator with 
covariant derivative diagram technique and matrix 
formalism were constructed. 
In particular, this formalism allows you to control the 
coefficients of the heat kernel, which is useful for calculations.
In this paper, a simple case is considered with abelian
connection in two-dimensional space.
This model allows us to give a mathematical 
description of operators and find relation between 
operators and generating functions of numbers.
\end{abstract}
\newpage
\section{Introduction}
This work is a consequence of the technique obtained \cite{1}  
in the study of combinatorial properties of the heat kernel for 
Laplace operator with covariant derivative. In short, 
the main result of paper can be formulated as follows.\\

\noindent Heat kernel $K(x,y,\tau)$ for d-dimensional ($d>1$)
Laplace operator with 
covariant derivative $\partial_{\mu}+\omega_{\mu}(x)$ 
can be represented as an asymptotic series 
$\sum\limits_{n=0}^{\infty}\tau^na_n(x,y)$, 
the coefficients of which are called 
Seeley-DeWitt coefficients. 
The calculation of these coefficients is
a time-consuming task that has been studied for many years \cite{15}. 
The paper proposes a technique that allows you to get an answer 
for an arbitrary coefficient (previously, similar methods were built
\cite{23}, but they have fundamental differences). 
The main elements of the technique are operators 
$A^{\mu}$, $B^{\mu}$, $S_l^{\mu}$ and $S_l^{\mathfrak{1}}$, 
which act on matrices with two rows and an arbitrary number of columns and 
defined by
\begin{equation*}
B^{\mu}
\begin{pmatrix}
\nu_1&\nu_2 & \ldots & \nu_n\\
k_1 & k_2 & \ldots & k_n
\end{pmatrix}
=
\begin{pmatrix}
\mu\nu_1&\nu_2 & \ldots & \nu_n\\
k_1+1 & k_2 & \ldots & k_n
\end{pmatrix}
+
\ldots+
\begin{pmatrix}
\nu_1&\nu_2 & \ldots & \mu\nu_n\\
k_1+1 & k_2+1 & \ldots & k_n+1
\end{pmatrix}\,,
\end{equation*}
\begin{equation*}
B^{\mu}
\begin{pmatrix}
\mathfrak{1}\\
k
\end{pmatrix}
=0\,,
\end{equation*}
\begin{multline*}
A^{\mu}
\begin{pmatrix}
\nu_1&\nu_2 & \ldots & \nu_n\\
k_1 & k_2 & \ldots & k_n
\end{pmatrix}
=
\begin{pmatrix}
\mu & \nu_1&\nu_2 & \ldots & \nu_n\\
k_1+1 & k_1 & k_2 & \ldots & k_n
\end{pmatrix}
+\ldots\\
\ldots+
\begin{pmatrix}
\nu_1&\nu_2 & \ldots & \mu & \nu_n\\
k_1+1 & k_2+1 & \ldots & k_n+1 &k_n
\end{pmatrix}
+
\begin{pmatrix}
\nu_1&\nu_2 & \ldots & \nu_n & \mu\\
k_1+1 & k_2+1 & \ldots & k_n+1 & 1
\end{pmatrix}\,,
\end{multline*}
\begin{equation*}
S_l^{\mu}
\begin{pmatrix}
\nu_1&\nu_2 & \ldots & \nu_n\\
k_1 & k_2 & \ldots & k_n
\end{pmatrix}
=
\begin{pmatrix}
\mu & \nu_1&\nu_2 & \ldots & \nu_n\\
l & k_1 & k_2 & \ldots & k_n
\end{pmatrix}\,,
\end{equation*}
where $\mu,\nu_1,\ldots,\nu_n\in\{1,\ldots,d\}$ and 
$l,k_1\ldots,k_n\in\mathbb{N}$. Anologeously 
$S_l^{\mathfrak{1}}$ can be introduced.
Also, an operation $\Upsilon$ was introduced, 
which maps a column to a tensor with some coefficient
according to rules
\begin{equation*}
\Upsilon
\begin{pmatrix}
\mu_{I_n}\nu\\
k
\end{pmatrix}
=
\frac{1}{k}
\sum\limits_{k=1}^{n}
\nabla_{\mu_{I_n\setminus k}}F_{\mu_k\nu}(x)\,\,,\,\,
\Upsilon
\begin{pmatrix}
\mathfrak{1}\\
k
\end{pmatrix}
=
\frac{1}{k}
\,,
\end{equation*}
where $k\in\mathbb{N}$, $I_n=\{1,\ldots,n\}$, 
$\nu,\mu_1,\ldots,\mu_n\in\{1,\ldots,d\}$,
$\mu_{I_n}=\mu_1\ldots\mu_n$ and
$F_{\mu\nu}(x)=\partial_{\mu}\omega_{\nu}(x)-
\partial_{\nu}\omega_{\mu}(x)+[\omega_{\mu}(x),\omega_{\nu}(x)]$.
It is assumed that the matrix is a set of columns.
Thus, the coefficients have the form:
\begin{equation}
\label{vv1}a_n(x,x)=
\Upsilon
\prod_{k=1}^{n}S^{\mathfrak{1}}_k(A^{\mu_k}+
B^{\mu_k})(A^{\mu_k}+B^{\mu_k})
\mathbb{1}\,,
\end{equation}
where the initial element $\mathbb{1}$ has proerties
\begin{equation*}
B^{\mu}\mathbb{1}=0\,\,,\,\,
A^{\mu}\mathbb{1}=
\begin{pmatrix}
\mu\\
1
\end{pmatrix}\,\,,\,\,
S^{\mu}_l\mathbb{1}=
\begin{pmatrix}
\mu\\
l
\end{pmatrix}\,\,,\,\,
S^{\mathfrak{1}}_l\mathbb{1}=
\begin{pmatrix}
\mathfrak{1}\\
l
\end{pmatrix}\,\,,\,\,
\Upsilon\mathbb{1}=1\,.
\end{equation*}
The main objects to study in this paper are the operators
$A$, $B$ and $\Upsilon$, the definition of which is given by formulas 
(\ref{oper1},
\ref{oper2} and \ref{oper3}). They are a simplified version of 
$A^{\mu}$, $B^{\mu}$ and $\Upsilon$.

\section{Problem formulation}
First of all, it is necessary to
make a few simplifications to give definitions of operators.
The paper will consider the case of action on $\mathbb{1}$ 
by operators with the same top indexes (
without loss of generality it can be $A^1$ and $B^1$).
It is seen
that the action of the operator $B^{\mu}$ on the first element of 
the column $\nu_1\ldots\nu_n$ is adding an additional index $\mu$.
It is proposed to consider a special case when 
the top index shows how many times the operator $B^1$ acted on 
the column.
Thus, using operators $B^{\mu}$, $A^{\mu}$, $\Upsilon$ 
and simplification mentioned above, definitions can be given:\\
\textbf{Definition 1:}
\begin{multline}
\label{oper2}
A
\begin{pmatrix}
s_1&s_2 & \ldots & s_n\\
k_1 & k_2 & \ldots & k_n
\end{pmatrix}
=
\begin{pmatrix}
0 & s_1&s_2 & \ldots & s_n\\
k_1+1 & k_1 & k_2 & \ldots & k_n
\end{pmatrix}
+\ldots\\
\ldots
+
\begin{pmatrix}
s_1&s_2 & \ldots & 0 & s_n\\
k_1+1 & k_2+1 & \ldots & k_n+1 &k_n
\end{pmatrix}
+
\begin{pmatrix}
s_1&s_2 & \ldots & s_n & 0\\
k_1+1 & k_2+1 & \ldots & k_n+1 & 1
\end{pmatrix}\,.
\end{multline}
\noindent\textbf{Definition 2:}
\begin{multline}
\label{oper1}
B
\begin{pmatrix}
s_1&s_2 & \ldots & s_n\\
k_1 & k_2 & \ldots & k_n
\end{pmatrix}
=
\begin{pmatrix}
s_1+1&s_2 & \ldots & s_n\\
k_1+1 & k_2 & \ldots & k_n
\end{pmatrix}
+\ldots\\
\ldots+
\begin{pmatrix}
s_1&s_2+1 & \ldots & s_n\\
k_1+1 & k_2+1 & \ldots & k_n
\end{pmatrix}
+
\begin{pmatrix}
s_1&s_2 & \ldots & s_n+1\\
k_1+1 & k_2+1 & \ldots & k_n+1
\end{pmatrix}\,.
\end{multline}
\noindent\textbf{Definition 3:}
\begin{equation}
\label{oper3}
\Upsilon
\begin{pmatrix}
s_1&s_2 & \ldots & s_n\\
k_1 & k_2 & \ldots & k_n
\end{pmatrix}
=\prod_{i=1}^n\left(\frac{s_i}{k_i}\right)\,,
\end{equation}
where $k_1,\ldots,k_n\in\mathbb{N}$ and $s_1,\ldots,s_n\in
\mathbb{N}\cup\{0\}$.
Also it is necessary to enter an element $\mathbb{1}$ 
that is defined by the following conditions:
\begin{equation}
\label{oper4}
B\mathbb{1}=0\,\,,\,\,
A\mathbb{1}=
\begin{pmatrix}
0\\
1
\end{pmatrix}\,\,,\,\,
\Upsilon\mathbb{1}=1\,.
\end{equation}
The main aim of this work is to give a mathematically correct definition of 
operators $A$, $B$ and $\Upsilon$ and to find a generating function 
for numbers
\begin{equation}
\label{oper5}
\varpi_n=\Upsilon(A+B)^n\mathbb{1}\,\,,\,\,n\geqslant0\,.
\end{equation}
\section{Auxiliary model}
\subsection{Motivation}
It is necessary to formulate some heuristic conditions, that will 
give rise to a convenient choice of 
connection components $\omega_{\mu}(x)$ and dimension $d$:
\begin{itemize}
 \item commutativity of connection components (due to dealing with numbers);
 \item the first connection component is zero ($\omega_1(x)=0$);
 \item $d=2$ (for simplicity).
\end{itemize}
Also it is useful to note, that the property $\left.\partial_x^n(xe^x)\right|_{x=0}=n$
can be used because the number in the first row shows how many 
times the operator $B$ has acted on the column.
\subsection{Two-dimensional model}
This set of conditions allows you to find the second connection component 
$\omega_2(x_1,x_2)$, where $x=(x_1,x_2)$. Also 
it is known from the construction of matrix formalism,
that action of $B$ is controlled by tension 
$F_{\mu\nu}(x)$ in such way, that
the last condition can be formulated as
\begin{equation}
\label{vs1}
x_2F_{21}(x_1,x_2)=x_1e^{x_1}\,.
\end{equation}
\begin{lemma}
Classical solution for (\ref{vs1}) 
with condition $\omega_2(0,x_2)=0$ when $x_2\neq0$ has the form
\begin{equation}
\label{vs3}
\omega_2(x_1,x_2)=\frac{1}{x_2}\left(e^{x_1}(1-x_1)-1\right)\,.
\end{equation}
\end{lemma}
\noindent In view of this, it is possible to write down a formal 
problem of finding a heat kernel:
\begin{equation*}
\begin{cases}
\left(\frac{\partial}{\partial\tau}+H\right)K(x,y;\tau)=0;\\
K(x,y;0)=\delta(x-y)\,,
\end{cases}
\end{equation*}
where
\begin{equation*}
\label{vs5}
H:=-\partial_{x_1}^2-\left(\partial_{x_2}+
\frac{1}{x_2}\left(e^{x_1}(1-x_1)-1\right)\right)^2\,.
\end{equation*}
\subsection{Path-ordered exponential}
In the case of commuting connection components path-ordered 
exponetial
\begin{equation*}
\Phi(x,y):=1+\sum\limits_{n=1}^{\infty}(-1)^n
\int\limits_{0}^{1}\int\limits_{0}^{1}\ldots\int\limits_{0}^{1}
ds_1\ldots ds_n
\frac{dz_1^{\nu_1}}{ds_1}\ldots\frac{dz_n^{\nu_n}}{ds_n}
\omega_{\nu_1}(z_1)\ldots \omega_{\nu_n}(z_n)
\end{equation*}
can be reduced to an exponential function.
\begin{lemma}
If $x=(x_1,x_2)$ and $y=(y_1,y_2)$, then
\begin{equation}
\label{vs6}
\Phi(x,y)=\exp\left\{\int\limits_{y_1}^{x_1}\frac{dt}{t}
(e^t(t-1)+1)\right\}\,.
\end{equation}
\end{lemma}
\noindent\textbf{Proof:} one can use answer (\ref{vs3})
and parametrization $z_{\mu}(s)=(1-s)y_{\mu}+sx_{\mu}$
\begin{equation}
-\int\limits_0^1ds\frac{dz_{\mu}(s)}{ds}\omega_{\mu}(z(s))=
-\int\limits_0^1ds\frac{dz_{2}(s)}{ds}\omega_{2}(z(s))=
\int\limits_{y_1}^{x_1}\frac{dt}{t}
(e^{t}(t-1)+1)\,.
\end{equation}
$\blacksquare$
\subsection{Generating function}
\begin{theorem}
\begin{equation}
\label{po1}
\Upsilon(A+B)^n\mathbb{1}=\left.\partial_x^n
\exp\left\{\int\limits_{0}^{x}\frac{dt}{t}
(e^t(t-1)+1)\right\}\right|_{x=0}\,.
\end{equation}
\end{theorem}
\noindent\textbf{Proof:} it is necessary to note that
\begin{equation*}
(\partial_{\mu}+\omega_{\mu}(x))\Phi(x,y)=
\int\limits_{0}^{1}dss\frac{dz_{\nu}(s)}{ds}\Phi(x,z(s))
F_{\nu\mu}(z(s))\Phi(z(s),y)\,,
\end{equation*}
which in considered case (\ref{vs1}) is reduced to
\begin{equation}
\label{po3}
\partial_{1}\Phi(x,0)=
\int\limits_{0}^{1}ds\Phi(x,z(s))
z_1(s)e^{z_1(s)}\Phi(z(s),y)\,.
\end{equation}
The proof follows from the theorem about differentiation of diagrams because:
\begin{itemize}
 \item the numbers of the upper row are obtained due to 
 the action of the derivative on factors of the form $ze^{z}$, 
 and the degrees of parameterization parameters standing to 
 the left are increased;
 \item the numbers of the lower row are obtained due to the action 
 of derivatives on the ordered exponentials, which leads to the appearance 
 of a new column and the change of degrees.
\end{itemize}
$\blacksquare$
\section{Algebraic structure}
The main motivation for the description of this combinatorics was 
the operator $A$, which maps a matrix with $n$ columns into a set 
of matrices with $n+1$ columns. At the same time 
matrices are equal to each other only when all their elements are the same.
All this suggests the introduction of vector space, comultiplication and other 
operations that will lead to bialgebras (see \cite{41}).
\subsection{Monoids}
For further work let us enter two sets
\begin{equation}
\label{osn2}
G_1=\mathbb{N}\cup\{0\}\,,
\end{equation}
\begin{equation}
\label{osn1}
G_2=\left\{
\frac{1}{q}:q\in\mathbb{N}\right\}\cup\{\infty\}\,,
\end{equation}
where infinity is introduced in the sense of the inverse 
element to zero with respect to the standard product of numbers. 
Also it is possible to introduce a binary operation $\star$ on $G_2$, 
which maps $\forall\,p,q\in G_2$ according to the following rule:
\begin{equation}
\label{osn3}
\star:(p,q)\longmapsto\left(\frac{1}{p}+
\frac{1}{q}\right)^{-1}\in G_2\,.
\end{equation}
Since $\star$ is assoceative and the element $\infty$ is a unit in
$G_2$ with respect to the operation then
\begin{lemma}
$G_1$ and $G_2$ are monoids with respect to operations 
$+$ and $\star$ respectively.
\end{lemma}
\noindent Using $G_1$ and $G_2$ it is possible to define two 
vector spaces $\mathbb{R}^{G_i}\,,\,i=1,2$, 
consisting of linear combinations of basis vectors 
$\mathbb{e}_q^i$, where $q\in G_i$, respectively.
\subsection{Bialgebras}
In order to construct a bialgebra, it is necessary to complete 
the vector space to the unital associative algebra and to the counital 
coassociative coalgebra with the necessary properties, that can be 
represented in the form of four commutative diagrams. 
On this account, four operations on $\mathbb{R}^{G_1}$ 
can be defined:\\
1) multiplication $\mu_1:\mathbb{R}^{G_1}\otimes\mathbb{R}^{G_1}
\longmapsto\mathbb{R}^{G_1}$, acting under the rule
\begin{equation}
\label{st1}
\mu_1(\mathbb{e}_p^1\otimes\mathbb{e}_q^1)=\mathbb{e}_{p+q}^1\,,\,
\forall\mathbb{e}_p^1\,,
\mathbb{e}_q^1\in\mathbb{R}^{G_2}\,;
\end{equation}
2) unit $\eta_1:\mathbb{R}\longmapsto\mathbb{R}^{G_1}$:
\begin{equation}
\label{st2}
\eta_1(1)=\mathbb{e}_{0}^1\,;
\end{equation}
3) comultiplication  $\Delta_1:\mathbb{R}^{G_1}
\longmapsto\mathbb{R}^{G_1}\otimes\mathbb{R}^{G_1}$, acting under the rule
\begin{equation}
\label{st3}
\Delta_1(\mathbb{e}_q^1)=\mathbb{e}_q^1\otimes\mathbb{e}_q^1\,,\,\forall
\mathbb{e}_q^1\in\mathbb{R}^{G_1}\,;
\end{equation}
4) counit $\varepsilon_1:\mathbb{R}^{G_1}\longmapsto\mathbb{R}$:
\begin{equation}
\label{st4}
\varepsilon_1(\mathbb{e}_{q}^1)=1\,,\,\forall
\mathbb{e}_q^1\in\mathbb{R}^{G_1}\,.
\end{equation}
Similarly operations $\mu_2$, $\eta_2$, $\Delta_2$ and 
$\varepsilon_2$ constructed for $\mathbb{R}^{G_2}$ 
by the replacement of the upper index at the basis
and by substitution $\star$ instead of $+$  and $\infty$ instead of $0$.
\begin{lemma} $(\mathbb{R}^{G_1},\mu_1,\eta_1)$ 
and $(\mathbb{R}^{G_2},\mu_2,\eta_2)$ are
unital, associative, commutative algebras.
\end{lemma}
\begin{lemma}
$(\mathbb{R}^{G_1},\Delta_1,\varepsilon_1)$ and 
$(\mathbb{R}^{G_2},\Delta_2,\varepsilon_2)$ are
counital, coassociative, cocommutative coalgebras.
\end{lemma}
\begin{lemma}
$(\mathbb{R}^{G_1},\mu_1,\eta_1,
\Delta_1,\varepsilon_1)$ and
$(\mathbb{R}^{G_2},\mu_2,\eta_2,
\Delta_2,\varepsilon_2)$ are bialgebras.
\end{lemma}
\begin{theorem}
$(V,\mu,\eta,
\Delta,\varepsilon)$ is bialgebra, where
\begin{equation}
\label{st5}
V=
\begin{pmatrix}
\mathbb{R}^{G_1}\\
\mathbb{R}^{G_2}
\end{pmatrix}\,\,,\,\,
\mu=
\begin{pmatrix}
\mu_1\\
\mu_2
\end{pmatrix}\,\,,\,\,
\eta=
\begin{pmatrix}
\eta_1\\
\eta_2
\end{pmatrix}\,\,,\,\,
\Delta=
\begin{pmatrix}
\Delta_1\\
\Delta_2
\end{pmatrix}\,\,,\,\,
\varepsilon=
\begin{pmatrix}
\varepsilon_1\\
\varepsilon_2
\end{pmatrix}\,\,.
\end{equation}
\end{theorem}

\subsection{Tensor algebra}
It is necessary to introduce maps $|\,\,\,|_1$ and $|\,\,\,|_2$ from 
$\mathbb{R}^{G_1}$ and $\mathbb{R}^{G_2}$ respectively to $\mathbb{R}$ 
operating under the following rules (city block distance \cite{43})  :
\begin{equation}
\label{ten1}
|\,\,\,|_1:\mathbb{R}^{G_1}\longmapsto\mathbb{R}:
\left|\sum\limits_{g\in G_1}\alpha_g\mathbb{e}_g^1\right|_1=
\sum\limits_{g\in G_1}|\alpha_g|g\,\,,\,\,\forall\alpha_g
\in\mathbb{R}\,,
\end{equation}
\begin{equation}
\label{ten2}
|\,\,\,|_2:\mathbb{R}^{G_2}\longmapsto\mathbb{R}:
\left|\beta_{\infty}\mathbb{e}_{\infty}^2+
\sum\limits_{g\in G_2\diagdown\{\infty\}}
\beta_g\mathbb{e}_g^2\right|_2=
\sum\limits_{g\in G_2\diagdown\{\infty\}}
|\beta_g|g\,\,,\,\,\forall\beta_g
\in\mathbb{R}\,.
\end{equation}
\begin{lemma} $|\,\,\,|_1$ and $|\,\,\,|_2$ are
seminorms on $\mathbb{R}^{G_1}$ and $\mathbb{R}^{G_2}$ respectively.
\end{lemma}
\noindent The objects introduced above 
can be generalized on $V$ according to the 
following formula
\begin{equation}
\label{ten3}
|\,\,\,|:V\longmapsto\mathbb{R}:
\left|
\sum\limits_{g_1\in G_1\,,\,g_2\in G_2}
\alpha_{g_1,g_2}\begin{pmatrix}
\mathbb{e}_{g_1}^1\\
\mathbb{e}_{g_2}^2
\end{pmatrix}
\right|=
\sum\limits_{g_1\in G_1\,,\,g_2\in G_2\diagdown\{\infty\}}
|\alpha_{g_1,g_2}|g_1g_2\,\,,\,\,\forall\alpha_{g_1,g_2}
\in\mathbb{R}\,.
\end{equation}
This is evident from the construction that $|\,\,\,|$ is seminorm on $V$. 
Because of it after introduction of tensor algebra
\begin{equation}
\label{ten4}
T(V)=\bigoplus\limits_{n=0}^{\infty}V^{\otimes k}\,,
\end{equation}
seminorm $|\,\,\,|$ can be expanded from $V$ to
$T(V)$ similar to the rule (\ref{ten3}) taking into account that on 
$\mathbb{R}$ it is equal to modulus. For 
$H=T(V)/\stackrel{L}{\sim}$,
where $L=\{v\in T(V):|v|=0\}$, $|\,\,\,|$ is the norm on $H$.

\subsection{Algebraic definition of operators}
It is possible to introduce on bialgebras $\mathbb{R}^{G_1}$ and
$\mathbb{R}^{G_2}$ two maps:
\begin{equation}
\label{sm1}
\theta_1:\mathbb{R}\longmapsto\mathbb{R}^{G_1}\,\,,\,\,
\theta_2:\mathbb{R}\longmapsto\mathbb{R}^{G_2}\,,
\end{equation}
acting according to the rules
\begin{equation}
\label{sm2}
\theta_1(1)=\mathbb{e}_1^1\,\,,\,\,
\theta_2(1)=\mathbb{e}_1^1\,.
\end{equation}
In this case operators $A$, $B$ and $\Upsilon$
take on quite a clear meaning
$\mathcal{A}$, $\mathcal{B}$ and $|\,\,\,|$ which can be formulated in
\begin{theorem} 
Operator $\mathcal{A}:T(V)\longmapsto T(V)$ according to the rule:
\begin{equation}
\label{sm3}
\mathcal{A}(1)=\begin{pmatrix}
\eta_1\\
\theta_2
\end{pmatrix}1=
\begin{pmatrix}
\mathbb{e}_{0}^1\\
\mathbb{e}_{1}^2
\end{pmatrix}\,,
\end{equation}
\begin{multline}
\label{sm4}
\mathcal{A}(v)=\sum\limits_{j=0}^{k-1}
\mu^{\otimes(j+1)}
\otimes
id^{\otimes(k-j)}
\circ
\left(
\begin{pmatrix}
\eta_1\\
\theta_2
\end{pmatrix}\otimes id\right)^{\otimes j}\otimes
\begin{pmatrix}
\eta_1\otimes\eta_1\otimes id\\
\theta_2\otimes\Delta_2
\end{pmatrix}\otimes
id^{\otimes(k-1-j)}v+\\+
\mu^{\otimes(k)}
\otimes id
\circ
\left(
\begin{pmatrix}
\eta_1\\
\theta_2
\end{pmatrix}\otimes id\right)^{\otimes k}\otimes
\begin{pmatrix}
\eta_1\\
\theta_2
\end{pmatrix}v\,\,,\,\,
\forall v\in V^{\otimes k}\,\,,\,\,\forall k\in\mathbb{N}\,.
\end{multline}
Operator $\mathcal{B}:T(V)\longmapsto T(V)$ according 
to the rule: $\mathcal{B}(1)=0$\,,
\begin{multline}
\label{sm5}
\mathcal{B}(v)=\sum\limits_{j=0}^{k-1}
\mu^{\otimes(j+1)}
\otimes
id^{\otimes(k-1-j)}
\circ
\left(
\begin{pmatrix}
\eta_1\\
\theta_2
\end{pmatrix}\otimes id\right)^{\otimes j}\otimes
\begin{pmatrix}
\theta_1\otimes id\\
\theta_2\otimes id
\end{pmatrix}\otimes
id^{\otimes(k-1-j)}v\,\,,\\
\forall v\in V^{\otimes k}\,\,,\,\,\forall k\in\mathbb{N}\,,
\end{multline}
Operator $\Upsilon:T(V)\longmapsto \mathbb{R}_+\cup\{0\}$:
\begin{equation}
\label{sm9}
\Upsilon(v)=|v|\,,\,\forall v\in T(V)\,.
\end{equation}
\end{theorem}
\noindent\textbf{Proof:} formulas (\ref{sm4}) and (\ref{sm5})
consist of blocks, so to prove the theorem it is enough to 
check the action on the basis vectors of $V$:
\begin{equation}
\label{sm6}
\mu
\circ
\begin{pmatrix}
\eta_1\\
\theta_2
\end{pmatrix}\otimes id
\begin{pmatrix}
\mathbb{e}_{g_1}^1\\
\mathbb{e}_{g_2}^2
\end{pmatrix}=
\mu
\begin{pmatrix}
\mathbb{e}_{0}^1&\mathbb{e}_{g_1}^1\\
\mathbb{e}_{1}^2&\mathbb{e}_{g_2}^2
\end{pmatrix}=
\begin{pmatrix}
\mathbb{e}_{g_1}^1\\
\mathbb{e}_{g_2\star 1}^2
\end{pmatrix}\,,\,
\forall g_1\in G_1\,,\,
\forall g_2\in G_2\,,
\end{equation}
\begin{equation}
\label{sm7}
\mu\otimes id
\circ
\begin{pmatrix}
\eta_1\otimes\eta_1\otimes id\\
\theta_2\otimes\Delta_2
\end{pmatrix}
\begin{pmatrix}
\mathbb{e}_{g_1}^1\\
\mathbb{e}_{g_2}^2
\end{pmatrix}=
\mu\otimes id
\begin{pmatrix}
\mathbb{e}_{0}^1&\mathbb{e}_{0}^1&\mathbb{e}_{g_1}^1\\
\mathbb{e}_{1}^2&\mathbb{e}_{g_2}^2&\mathbb{e}_{g_2}^2
\end{pmatrix}=
\begin{pmatrix}
\mathbb{e}_{0}^1&\mathbb{e}_{g_1}^1\\
\mathbb{e}_{g_2\star 1}^2&\mathbb{e}_{g_2}^2
\end{pmatrix}\,,\,
\forall g_1\in G_1\,,\,
\forall g_2\in G_2\,,
\end{equation}
\begin{equation}
\label{sm8}
\mu
\circ
\begin{pmatrix}
\theta_1\otimes id\\
\theta_2\otimes id
\end{pmatrix}
\begin{pmatrix}
\mathbb{e}_{g_1}^1\\
\mathbb{e}_{g_2}^2
\end{pmatrix}=
\mu
\begin{pmatrix}
\mathbb{e}_{1}^1&\mathbb{e}_{g_1}^1\\
\mathbb{e}_{1}^2&\mathbb{e}_{g_2}^2
\end{pmatrix}=
\begin{pmatrix}
\mathbb{e}_{g_1+1}^1\\
\mathbb{e}_{g_2\star 1}^2
\end{pmatrix}\,,\,
\forall g_1\in G_1\,,\,
\forall g_2\in G_2\,.
\end{equation}
$\blacksquare$

\section{Generalization}
The resulting combinatorics was based on the assumption
$x_2F_{21}(x_1,x_2)=x_1e^{x_1}$, which controlled the action 
of the operator $B$ by the rule (\ref{oper1}). However, this 
assumption can be abandoned, thus expanding the 
class of operators.
\subsection{Generating function:}
To this end, the equality of the more general situation 
should be considered
\begin{equation}
\label{ob1}
x_2F_{21}(x_1,x_2)=f(x_1)\,,
\end{equation}
where the decomposition of the function into a series in 
the neighborhood of zero is given by the formula
\begin{equation}
\label{ob2}
f(x_1)=\sum\limits_{k=0}^{\infty}\frac{b_k}{k!}x_1^k\,.
\end{equation}
Under these assumptions, the generating function is written explicitly:
\begin{equation}
\label{ob3}
\Phi(x_1,0)=\exp\left\{
\int\limits_{0}^{x_1}\frac{dt}{t}\int\limits_{0}^{t}
dsf(s)
\right\}\,.
\end{equation}
\subsection{Algebraic structure:}
In the case (\ref{ob1}), when an algebraic structure is 
discribed, only the definition of seminorm $|\,\,\,|_1$, 
the kernel of the seminorm and normalization of 
the basis vectors, should be changed.
Formally, this transformation is:
\begin{equation}
\label{ob4}
|\mathbb{e}_g^1|_1=g\,,\,\forall g\in G_1\longrightarrow
|\mathbb{e}_g^1|_1=b_g\,,\,\forall g\in G_1\,,
\end{equation}
\begin{equation}
\label{ob5}
\mathtt{Ker}(|\,\,\,|_1)=\{\mathbb{e}_0^1\}\longrightarrow
\mathtt{Ker}(|\,\,\,|_1)=\{\mathbb{e}_g^1
\,,\,\forall g\in G_1:b_g=0\}\,.
\end{equation}
\begin{theorem}
The transition from the model with condition
$x_2F_{21}(x_1,x_2)=x_1e^{x_1}$ to the model with condition
$x_2F_{21}(x_1,x_2)=f(x_1)$ can be done by 
change of generatin function to (\ref{ob3}) and seminorm to
(\ref{ob4}) and (\ref{ob5}).
\end{theorem}
\noindent In the case when the function $f(x)$ has a negative 
Taylor coefficient,
the map $|\,\,\,|_1$ loses the sense of seminorm.
\subsection{Inverse problem}
The reverse process is also possible: 
the function $f(x_1)$ can be constructed on the basis 
of an ordered exponential $\Phi(x_1,0)$.
\begin{lemma}
\begin{equation}
\label{ob6}
f(x_1)=\frac{\Phi(x_1,0)\Phi'(x_1,0)+
x_1(\Phi(x_1,0)\Phi''(x_1,0)-
(\Phi'(x_1,0))^2)}{(\Phi(x_1,0))^2}\,.
\end{equation}
\end{lemma}
\noindent\textbf{Proof:}
\begin{equation*}
\omega_2(x_1,x_2)=-\frac{x_1}{x_2}\partial_{x_1}\left(
\ln(\Phi(x_1,0))\right)\,,
\end{equation*}
\begin{equation*}
x_2F_{21}(x_1,x_2)=-x_2\partial_{x_1}\omega_2(x_1,x_2)\,.
\end{equation*}
$\blacksquare$\\
\noindent 
It should be noted that the generating function $\Phi(x_1,0)$ 
has a special property, namely $\Phi(0,0)=1$.
Naturally, not every generating function has this property,
so the formula (\ref{ob6}) is not applicable for all 
generating functions due to problem \cite{49}. 
However, there are several obvious solutions:
\begin{theorem} If $F(x_1)$ is
the generating function of numbers such that
$F(0)\neq0$, then the function
$f(x_1)$ can be constructed by the formula (\ref{ob6}) for
\begin{equation}
\label{ob9}
\Phi(x_1,0)=\frac{F(x_1)}{F(0)}\,,
\end{equation}
which leads to the replacement of the initial data 
$1\mapsto F(0)$ in (\ref{oper5})
\begin{equation}
\label{ob10}
\varpi_n=\Upsilon(\mathcal{A}+\mathcal{B})^nF(0)\,.
\end{equation}
\end{theorem}
\begin{theorem}
If $F(x_1)$ is arbitrary generating function of numbers,
then the function 
$f(x_1)$ can be constructed by the formula (\ref{ob6}) for
\begin{equation}
\label{ob11}
\Phi(x_1,0)=F(x_1)-F(0)+1\,,
\end{equation}
which leads to the answer in the form
\begin{equation}
\label{ob12}
\varpi_n=\Upsilon(\mathcal{A}+\mathcal{B})^n1+(F(0)-1)\delta_{n0}\,.
\end{equation}
\end{theorem}
\subsection{Examples}
\begin{center}
\begin{tabular}{|c|c|c|}
\hline
Numbers & $F(x)$ & $f(x)$\\
\hline
Catalan numbers \cite{44}&$\frac{1-\sqrt{1-4x}}{2x}
$&$
\frac{1}{(1-4x)^{3/2}}$\\
\hline
Bell numbers \cite{46}&$e^{e^x-1}$&$e^x(x+1)$\\
\hline
Binomial expansion&
$(1+x)^{\alpha}\,,\forall\alpha\in\mathbb{C}$&
$\frac{\alpha}{(1+x)^2}\,,\forall\alpha\in\mathbb{C}$\\
\hline
Exponential numbers \cite{45}&
$e^{sin(x)}$&$\cos(x)-x\sin(x)$\\
\hline
\end{tabular}
\end{center}

\section{Corollaries and comments}
\subsection{Operator function}
The numbers of view
$\Upsilon\phi(\mathcal{A},\mathcal{B})1$ are
of greatest interest of this section,
where the function is monomial of operators. The special 
case $\phi(\mathcal{A},\mathcal{B})=(\mathcal{A}+\mathcal{B})^n$, 
where $n\in\mathbb{N}\cup\{0\}$, has been studied above. Of course 
the main aim is to find
\begin{equation*}
\Upsilon\prod\limits_{k=1}^{n}\Lambda_k^{\sigma_k}1\,,
\end{equation*}
where $n\in\mathbb{N}$, $\Lambda_k\in\{\mathcal{A},\mathcal{B}\}$ and
$\sigma_k\in\mathbb{N}\cup\{0\}$ for $k\in\{1,\ldots,n\}$. 
Using diagram technique it is very easy to obtain, that 
operator $\mathcal{A}$ acts only on ordered exponentials (lines) 
while operator $\mathcal{B}$ differentiates all functions (circles) 
except exponentials. 
But due to the commutativity of connection components $\omega_{\mu}(x)$, 
the ordered exponential can be taken out from 
expression and can be done the factor $\Phi(x,0)$.
In this case, the operator $\mathcal{A}$ is operator of 
multiplication on function $\partial_x\ln\Phi(x,0)$.
Therefore the operator $\mathcal{B}$ acts like a derivative with 
connection $-\ln'_x\Phi(x,0)$.
\begin{theorem}
If 
$n\in\mathbb{N}$, $\Lambda_k\in\{\mathcal{A},\mathcal{B}\}$ and
$\sigma_k\in\mathbb{N}\cup\{0\}$ for $k\in\{1,\ldots,n\}$, and
$$\phi(\mathcal{A},\mathcal{B})=
\prod\limits_{k=1}^{n}\Lambda_k^{\sigma_k}\,,$$
then
\begin{equation*}
\Upsilon\phi(\mathcal{A},\mathcal{B})1
=\left.
\phi(\ln'_x\Phi(x,0),\partial_x-\ln'_x\Phi(x,0))
\Phi(x,0)\right|_{x=0}\,.
\end{equation*}
\end{theorem}
\subsection{Estimates for ordered exponential}
Using numbers $\varpi_{n}$, 
local estimates for derivatives of an ordered exponential 
can be written:
\begin{theorem}
If dimension is equal to $d$, 
$\nabla_{\mu}=\partial_{\mu}+\omega_{\mu}(x)$, $\Phi(x,y)$ is 
ordered exponential constructed by connection components 
$\omega_{\mu}(x)$ and in some neighborhood $U_{\delta}(x_0)$ of 
some point $x_0\in\mathbb{R}^d$ $\exists C>0:$ the inequality 
for $x\in U_{\delta}(x_0)$
\begin{equation}
\label{ko2}
|\nabla_{\mu_1}\cdot\ldots\cdot
\nabla_{\mu_{k-2}}F_{\mu_{k-1}\mu_k}(x)|
\leqslant C^k\,,\,\forall k\in\mathbb{N}\,,\,
\forall \mu_1,\ldots,\mu_k\in\{1,\ldots,d\}\,,
\end{equation}
is valid, then
\begin{equation}
\label{ko3}
\left|\left.\prod_{k=1}^{n}\nabla^{\mu_k}\nabla_{\mu_k}\Phi(x,y)
\right|_{y=x}\right|\leqslant d^nC^{2n}\varpi_{2n}\,,\,
\forall n\in\mathbb{N}\,,
\end{equation}
\begin{equation}
\label{ko4}
\left|\left.\prod_{k=1}^{n}
\int\limits_0^1dss^{k-1}s^{(x-y)^{\rho}\partial_{\rho}}
\nabla^{\mu_k}\nabla_{\mu_k}\Phi(x,y)
\right|_{y=x}\right|\leqslant d^nC^{2n}
\frac{\varpi_{2n}(n-1)!}{(2n-1)!}\,,\,
\forall n\in\mathbb{N}\,.
\end{equation}
\end{theorem}
\subsection{General remarks}
1) Returning to the formula (\ref{vv1}), it is very easy to see that
analogously
operator $S^k_l$ can be defined on $T(V)$ by the formula:
\begin{multline}
\label{ko1}
\mathcal{S}^k_lv=
\left(\mu\otimes id^{\otimes n}\circ
\begin{pmatrix}
\eta_1\\
\theta_2
\end{pmatrix}\otimes id^{\otimes n}\right)^{\circ l}\circ
\left(\mu\otimes id^{\otimes n}\circ
\begin{pmatrix}
\theta_1\\
\eta_2
\end{pmatrix}\otimes id^{\otimes n}\right)^{\circ k}\circ\\\circ
\begin{pmatrix}
\eta_1\\
\eta_2
\end{pmatrix}\otimes id^{\otimes n}v\,,\,
\forall v\in V^{\otimes n}\,,\,\forall n\in\mathbb{N}\cup\{0\}\,.
\end{multline}
2) The paper deals with elementary case. However, the general model
\begin{itemize}
 \item dimension $d>2$;
 \item non-commuting connections,
\end{itemize}
can be constructed by using matrix formalism of heat kernel.

\section{Acknowledgements}
This research is supported by a grant from 
the Russian Science Foundation (Project No. 14-11-00598).

\end{document}